\documentclass[aps, prd, twocolumn,
showpacs, showkeys, preprintnumbers, nofootinbib, superscriptaddress]{revtex4-1}

\usepackage{amssymb,amsbsy,amsmath,amsfonts}
\usepackage{slashed}
\usepackage{bm}
\usepackage{times}

\usepackage[usenames, dvipsnames]{color}

\usepackage{graphicx}
\graphicspath{{./}{./img/}{./fig/}{./image/}{./figure/}{./picture/}}
\usepackage{epstopdf}

\usepackage{multirow}
\usepackage{tabularx}
\usepackage{longtable}

\usepackage{hyperref}
\hypersetup{
  colorlinks=true,        
  linkcolor=blue,         
  citecolor=cyan,         
  urlcolor=red,           
}
\begin{document}

\title{Pion-nucleon sigma term revisited in covariant baryon chiral perturbation theory}

\author{Xiu-Lei Ren}
\affiliation{State Key Laboratory of Nuclear Physics and Technology, School of Physics, Peking University, Beijing 100871, China}
\affiliation{Institut f\"{u}r Theoretische Physik II, Ruhr-Universit\"{a}t Bochum, D-44780 Bochum, Germany}

\author{Xi-Zhe Ling}
\affiliation{School of Physics and
Nuclear Energy Engineering and International Research Center for Nuclei and Particles in the Cosmos, Beihang University, Beijing 100191, China}

\author{Li-Sheng Geng}
\email[E-mail: ]{lisheng.geng@buaa.edu.cn}
\affiliation{School of Physics and
Nuclear Energy Engineering and International Research Center for Nuclei and Particles in the Cosmos, Beihang University,
Beijing 100191, China}
\affiliation{Beijing Key Laboratory of Advanced Nuclear Materials and Physics, Beihang University, Beijing 100191, China}

\begin{abstract}
We study the latest $N_f=2+1+1$ and $N_f=2$ ETMC lattice QCD simulations of the nucleon masses and extract the pion-nucleon sigma term utilizing the
Feynman-Hellmann theorem in SU(2) baryon chiral perturbation theory with the extended-on-mass-shell scheme. We find that the lattice QCD data can be described
quite well already at the next-to-next-to-leading order. The overall picture remains essentially the same at the next-to-next-to-next-to-leading order. Our final result is
$\sigma_{\pi N}=50.2(1.2)(2.0)$ MeV, or equivalently, $f_{u/d}^N=0.0535(13)(21)$, where the first uncertainty is statistical and second is theoretical originated from chiral truncations,
 which is in agreement with that determined previously from the $N_f=2+1$  and $N_f=2$ lattice QCD data and that
determined by the Cheng-Dashen theorem.  In addition, we
show that the inclusion of the virtual $\Delta(1232)$ does not change qualitatively our results. 
\end{abstract}


\date{\today}

\maketitle

\section{Introduction} 
In recent years, the pion-nucleon sigma term has attracted much attention, partly because of its
role in predicting the cross section of certain candidate dark matter particles interacting with the nucleons~\cite{Bottino:1999ei}.   Historically, a ``canonical value'' of the pion-nucleon sigma term $\sigma_{\pi N}=m_l\langle N|\bar{u}u+\bar{d}d|N\rangle\sim 45$ MeV was derived in Ref.~\cite{Gasser:1990ce} from the pion-nucleon scattering data. 
Later, an updated analysis of $\pi N$ scattering yielded  a larger value $\sigma_{\pi N}=64(8)$ MeV~\cite{Pavan:2001wz}. In the past few years,  several phenomenological studies of pion-nucleon scattering using chiral perturbation theory (ChPT) and/or Roy-Steiner equations,  e.g.~Refs.~\cite{Alarcon:2011zs,Chen:2012nx,Yao:2016vbz,Hoferichter:2015dsa,Hoferichter:2015hva}, have derived a $\sigma_{\pi N}$ around $60$ MeV. In the meantime, the pion-nucleon sigma term has also been extensively studied in lattice quantum chromodynamics (lattice QCD) by either computing three-point (the direct method)~\cite{Bali:2011ks,Dinter:2012tt,Yang:2015uis,Abdel-Rehim:2016won,Bali:2016lvx} or two-point correlation functions (the so-called spectrum method)~\cite{Ohki:2008ff,Young:2009zb,Durr:2011mp,Horsley:2011wr,Semke:2012gs,Shanahan:2012wh,Ren:2012aj,Alvarez-Ruso:2013fza,Lutz:2014oxa,Ren:2014vea, Alexandrou:2014sha, Durr:2015dna}. 
Due to the many systematic and statistical uncertainties inherent in these studies, no consensus has been reached on the precise value of the pion-nucleon sigma term, although several recent studies seem to prefer a small value$\sim 40$ MeV~\cite{Yang:2015uis,Durr:2015dna,Abdel-Rehim:2016won,Bali:2016lvx}. Apparently, there exists a tension between the pion-nucleon sigma term determined from the phenomenological studies and that from the lattice QCD simulations. 

As stressed in Ref.~\cite{Ren:2014vea}, two key factors are  important in a reliable and accurate determination of the pion-nucleon sigma term using the lattice nucleon mass data with the spectrum method, i.e., lattice QCD simulations with
various setups and configurations and  a proper formulation to parameterize the pion-mass dependence of the nucleon mass. 
For the later, baryon chiral perturbation theory (BChPT), an effective field theory of low-energy QCD, provides a model-independent framework to study the pion-mass dependence of the nucleon mass.  
In the last few years,  the European Twisted Mass Collaboration (ETMC) has performed several lattice QCD studies to extract the nucleon mass
 with the $N_f=2$ \cite{Alexandrou:2009qu, Alexandrou:2017xwd} and $N_f=2+1+1$~\cite{Alexandrou:2014sha}  twisted mass fermions. 
Since the dynamical strange and charm quarks  have minor impact on the ETMC nucleon masses, 
in a recent work, Alexandrou {\it et al.} (ETMC) performed a combined fit to the 17 sets of the $N_f=2+1+1$ nucleon masses
 and one $N_f=2$ physical ensemble  using SU(2) BChPT~\footnote{In principle, the twisted-mass ChPT~\cite{Sharpe:2004ny,WalkerLoud:2005bt} is more suitable for the analysis of the ETMC data.}, and predicted 
a pion-nucleon sigma term $ 64.9(1.5) (13.2)$ MeV~\cite{Alexandrou:2017xwd}. 
This value is much larger than that obtained from the direct method with the ensemble at the physical point by the same collaboration, $\sigma_{\pi N}=37.2(2.6)^{(4.7)}_{(2.9)}$~\cite{Abdel-Rehim:2016won}. 
However, ones should note that the large $\sigma_{\pi N}$ of Ref.~\cite{Alexandrou:2017xwd} was obtained in the
spectrum method using the heavy baryon (HB) chiral perturbation theory, which is known to perform sometimes badly  in terms of convergence (see, e.g., Ref.~\cite{Pascalutsa;2004, MartinCamalich:2010fp}). Particularly, it was shown in Ref.~\cite{Alexandrou:2017xwd} that at next-to-next-to-leading order (NNLO) the best fit yields a $\chi^2/\mathrm{d.o.f.} \approx1.6$ while only at ``next-to-next-to-next-to leading order (N$^3$LO)''~\footnote{One should note that this is not a complete N$^3$LO study in HB ChPT, since the contributions from the $\mathcal{O}(p^4)$ tadpole and mass-insertion loop diagrams  were not included.}, a $\chi^2/\mathrm{d.o.f.} \approx1.1$ can be achieved. 

Since the determination of the pion-nucleon sigma term via the Feynman-Hellmann theorem is sensitive to the extracted pion-mass dependence of the nucleon mass from the lattice QCD data, a better description of the ETMC data is needed.  Therefore, it is timely and worthy to reanalyze the same lattice QCD data as Ref.~\cite{Alexandrou:2017xwd} using  covariant baryon chiral perturbation theory with the extended-on-mass-shell (EOMS) scheme~\cite{Gegelia:1999gf}, which has
shown a number of both formal and practical advantages and has solved a number of long-existing puzzles in the one-baryon sector~\cite{Geng:2013xn}. Furthermore, the applications of the EOMS BChPT in the studies of the lattice QCD octet baryon masses
 turn out to be very successful as well~\cite{Ren:2012aj,Ren:2013dzt,Ren:2013oaa,Alvarez-Ruso:2013fza}.~\footnote{It has been extended to heavy flavor sectors in recent years, see, e.g., Refs.~\cite{ Geng:2010vw,Altenbuchinger:2013vwa,Sun:2016wzh,Yao:2018ifh}.}
Therefore, in this work, we employ the two-flavor covariant BChPT to calculate the nucleon mass up to N$^3$LO. 
It is shown that we can achieve a better description of  the 18 sets of ETMC data, i.e. $\chi^2/\mathrm{d.o.f.}  \leq 1.0$, in comparison with the study in the HB scheme~\cite{Alexandrou:2017xwd}. With the obtained LECs, we predict a pion-nucleon sigma term, $\sigma_{\pi N}=50.2(1.2)(2.2)$ MeV, using the Feynman-Hellmann theorem. 
 
 This paper is organized as follows. In Section II, we briefly summarize the theoretical ingredients needed to analyze the ETMC lattice QCD data. In Section III, we
 perform fits to them following the strategy of Ref.~\cite{Alexandrou:2017xwd} and predict the pion-nucleon
sigma term using  the Feynman-Hellmann theorem. The so-obtained low-energy constants (LECs) are
then used to calculate the scattering length as well as the pion-nucleon sigma term with the Cheng-Dashen theorem. In Section IV, a short summary is given. 
 
\section{Theoretical framework}

The nucleon mass  has been calculated up to $\mathcal{O}(p^4)$ both in the two-flavor sector~\cite{Alvarez-Ruso:2013fza} and in the three-flavor sector~\cite{Ren:2012aj} in covariant BChPT with the EOMS scheme.  
To make the present work self-consistent, we spell out  the nucleon mass up to $\mathcal{O}(p^4)$, which in the isospin symmetric limit reads
\begin{eqnarray} \label{Eq:nucleonmass}
 m_N &=& m_0   - 4c_1m_\pi^2 +  \alpha m_\pi^4   + \frac{3c_2 m_\pi^4}{128\pi^2f_\pi^2} \nonumber\\
  &&  -\frac{3}{64\pi^2f_\pi^2}(8c_1-c_2-4c_3)  m_\pi^4 \left(1+\log\frac{\mu^2}{m_\pi^2}\right) \nonumber\\ 
  && + \frac{3g_A^2}{4(4\pi f_\pi)^2}  \left[  H_N^{(3)}(m_0,~m_\pi,~\mu) \right. \nonumber\\
  &&\quad +  \left. H_N^{(4)}(m_0,~(-4c_1m_\pi^2),~m_\pi,~\mu ) \right],
\end{eqnarray}
where $f_\pi$ is the pion decay constant in the chiral limit, and $g_A$ is
the axial coupling. 
There are four LECs, $c_1$, $c_2$, $c_3$, and $\alpha$. 
The two loop functions, $H_N^{(3)}$ and $H_N^{(4)}$, are the contributions of the $\mathcal{O}(p^3)$ and $\mathcal{O}(p^4)$ one-loop diagrams with the power-counting breaking terms subtracted~\cite{Alvarez-Ruso:2013fza,Ren:2012aj}
\begin{eqnarray}
  H_N^{(3)} &=& -\frac{2m_\pi^3}{m_0}
               \left[ \frac{m_\pi}{2} \log \frac{m_\pi^2}{m_0^2} +  \sqrt{4m_0^2-m_\pi^2} \right. \times \nonumber\\
            && \left( \arctan\frac{m_\pi}{\sqrt{4m_0^2-m_\pi^2}} \right.\nonumber\\
               && \left. \left. - \arctan\frac{m_\pi^2-2m_0^2}{m_\pi
    \sqrt{4m_0^2-m_\pi^2}} \right) \right],
\end{eqnarray} 
\begin{eqnarray}
   H_{N}^{(4)}&=& \frac{2m_\pi^3}{m_0^2\sqrt{4m_0^2-m_\pi^2}} (4c_1m_\pi^4)
   \arccos\frac{m_\pi}{2m_0} \nonumber \\
 &&-m_\pi^2 \left[\frac{4c_1m_\pi^4}{m_0^2}\log\frac{m_\pi^2}{m_0^2}-8c_1m_\pi^2 \log\frac{m_0^2}{\mu^2}\right],
 \end{eqnarray}
which are calculated in the dimensional regularization scheme with the renormalization scale $\mu$. 
Following Ref.~\cite{Ren:2016aeo}, we take $f_\pi =0.0871$ GeV, $g_A=1.267$, and $\mu=1.0$ GeV in our numerical study, unless otherwise specified.

\begin{table}[b]
\caption{ Eighteen sets of the $N_f=2+1+1$ and one $N_f=2$  ETMC data of Ref.~\cite{Alexandrou:2017xwd}.}
\begin{tabular}{c|c|c|c|c|c}  
\hline\hline
Set No.&Volume&Statistics&$a\mu_l$&$am_\pi$&$am_N$\\
\hline\hline  
\multicolumn{6}{c}{$N_f=2+1+1, \beta=1.90$}\\ \hline
  1&\multirow{3}{*}{$32^3\times64$} &2960&0.0030&0.1240&0.5239(87)\\  
   2&&6224&0.0040&0.1414&0.5192(112)\\ 
   3&&1548&0.0050&0.1580&0.5422(62)\\
    \hline
4&\multirow{4}{*}{$24^3\times48$}&8368&0.0400&0.1449&0.5414(84)\\
    5&&7664&0.0060&0.1728&0.5722(48)\\
    6&&7184&0.0080&0.1988&0.5898(50)\\
    7&&8016&0.0100&0.2229&0.6206(43)\\
    \hline
8&$20^3\times48$&2468&0.0040&0.1493&0.5499(195)\\
 \hline\hline
\multicolumn{6}{c}{$N_f=2+1+1, \beta=1.95$}\\ \hline
9&\multirow{4}{*}{$32^3\times64$}&2892&0.0025&0.1068&0.4470(59)\\
    10&&4204&0.0035&0.1260&0.4784(48)\\
    11&&18576&0.0055&0.1552&0.5031(16)\\
    12&&2084&0.0075&0.1802&0.5330(42)\\
\hline
13&$24^3\times48$&937&0.0085&0.1940&0.5416(50)\\
 \hline\hline
\multicolumn{6}{c}{$N_f=2+1+1, \beta=2.10$}\\ \hline
14&\multirow{3}{*}{$48^3\times96$}&2424&0.0015&0.0698&0.3380(41)\\ 
    15&&744&0.0020&0.0805&0.3514(70)\\  
    16&&904&0.0030&0.0978&0.3618(68)\\   
    \hline
17&$32^3\times64$&7620&0.0045&0.1209&0.3944(26)\\  
 \hline\hline
\multicolumn{6}{c}{$N_f=2, \beta=2.10, c_{sw}=1.57551$}\\ \hline
18&$48^3\times96$&861200&0.0009&0.0621&0.4436(11)\\
\hline\hline
\end{tabular}
\label{Tab:LQCDdata}
\end{table}

\begin{table*}[t]
\caption{ Fitted LECs of the $\mathcal{O}(p^3)$ and $\mathcal{O}(p^4)$  EOMS BChPT, as well as the predicted $\sigma_{\pi N}$. The numbers in the parentheses are the
statistical uncertainties at the 68.3\% confidence level.}
\centering 
\begin{tabular}{c|c|ccccc|c}  
\hline\hline
  & $\chi^2/\mathrm{d.o.f.}$  &$m_0~(\mathrm{GeV})$ & $c_1~(\mathrm{GeV}^{-1})$ & $\alpha~(\mathrm{GeV}^{-3})$  & $c_2~(\mathrm{GeV}^{-3})$ & $c_3~(\mathrm{GeV}^{-3})$ & $\sigma_{\pi N}~(\mathrm{MeV})$ \\
  \hline  
  $\mathcal{O}(p^3)$& 0.87 & $0.882\pm 0.002 $ & $ -0.95\pm 0.02 $ & -- & -- & -- & $50.2\pm 1.2$ \\
 $\mathcal{O}(p^4)$  & 0.75 &  $0.879\pm 0.010$  &  $-1.03 \pm 0.20 $ & $7.31\pm 9.43$ & $-2.34\pm 4.14$ & $-2.67\pm 1.60$ & $52.2\pm 6.6$ \\
\hline\hline
\end{tabular}
\label{Tab:LECs-sigma}
\end{table*}

In principle, the four LECs ($c_i$ and $\alpha$) can be calculated directly from QCD. However, because of  the nonperturbative nature of QCD at low energies,  one usually determines their value by performing a least-square fit to the lattice QCD nucleon masses and/or experimental data.
 It was shown in Refs.~\cite{Ren:2012aj,Lutz:2014oxa}  that finite volume corrections need to be taken into account, particularly for the $m_\pi L<4$ ensembles, in order to describe the lattice QCD data with a $\chi^2\approx1.0$.
  In the present case, since some of the ETMC results
are obtained with $m_\pi L<4$, we take the finite volume corrections  into account up to $\mathcal{O}(p^4)$, which read
\begin{eqnarray}
  \delta m_N &=& \frac{3g_A^2}{4f_\pi^2} \left( \delta H_{N}^{(3)}  + \delta H_{N}^{(4)} \right)
                 + \frac{3}{2f_\pi^2}  \left[ 2c_1 m_\pi^2 \delta_{1/2}(m_\pi^2) \right.\nonumber\\
   && \left. - c_2 \delta_{-1/2}(m_\pi^2)
  - c_3 m_\pi^2  \delta_{1/2}(m_\pi^2) \right],
\end{eqnarray}
with \begin{eqnarray}
\delta_{r}(\mathcal{M}^2) &=& \frac{2^{-1/2-r}(\sqrt{\mathcal{M}^2})^{3-2r}}{\pi^{3/2}\Gamma(r)}  \times \nonumber\\
&& \sum_{\vec{n}\neq0}(L\sqrt{\mathcal{M}^2}|\vec{n}|)^{-3/2+r}K_{3/2-r}(L\sqrt{\mathcal{M}^2}|\vec{n}|),
\end{eqnarray}
where $K_n(z)$ is the modified Bessel function of the second kind, and
\begin{equation}
  \sum_{\vec{n}\neq0}\equiv
  \sum_{n_x=-\infty}^{\infty}\sum_{n_y=-\infty}^{\infty}\sum_{n_z=-\infty}^{\infty}(1-\delta(|\vec{n}|,0)),
\end{equation}
with $\vec{n}=(n_x,n_y,n_z)$. The finite volume correction of the one-loop diagrams, $\delta H_N^{(3)}$ and $\delta H_N^{(4)}$, are
calculated in  Refs.~\cite{Geng:2011wq,Ren:2012aj} and read 
\begin{eqnarray}
 \delta H_N^{(3)} &=& - \int_0^1 dx \left[ \frac{1}{2} m_0 (2x+1) \delta_{1/2}(\mathcal{M}_N^2) \right. \nonumber\\
                  && \left. -\frac{1}{4}m_0\left(m_0^2 x^3 +\mathcal{M}_N^2(x+2)\right)\delta_{3/2}(\mathcal{M}_N^2)
                     \right],
\end{eqnarray}
and
\begin{eqnarray}
 \delta H_N^{(4)} & = & \int_0^1 dx \left\{ -\frac{1}{2}\delta_{1/2}(\mathcal{M}_N^2)(2x+1)m_N^{(2)}\right.\nonumber \\
 && -\frac{1}{4}\delta_{3/2}(\mathcal{M}_N^2)\left[m_\pi^2 (x-1)(x+2)m_N^{(2)}\right. \nonumber\\
&& \left. \quad  -2xm_0^2 (5x^2+4x)m_N^{(2)}\right]\nonumber \\
&& \left. -\frac{1}{4}\delta_{5/2}(\mathcal{M}_N^2)\left[ 6m_0^4(x+1)x^4 m_N^{(2)} \right.\right. \nonumber \\
&&\quad \left.\left.  -3m_0^2 m_\pi^2x^2(x-1)(x+2)m_N^{(2)}\right]\right\},
\end{eqnarray}
with the leading order correction to the nucleon self-energy $m_N^{(2)}=-4c_1 m_\pi^2$ and $\mathcal{M}_N^2=x^2m_0^2+(1-x)m_\pi^2-i\epsilon$.
 
Once we obtain the nucleon mass, the pion-nucleon sigma term can be predicted utilizing the Feynman-Hellmann theorem~\cite{Feynman:1939zza}, which dictates that in the isospin symmetric limit the $\sigma_{\pi N}$ can be calculated from the light quark mass or equivalently the pion mass dependence of the nucleon mass, $m_N$, in the following way
\begin{eqnarray}
  \sigma_{\pi N} &=& m_l\langle N|\bar{u}u+\bar{d}d|N\rangle \equiv m_l\frac{\partial m_N}{\partial m_l},
\end{eqnarray}
where leading order ChPT has been used to relate the light quark mass with the pion mass.~\footnote{We have checked that using 
the next-to-leading order ChPT instead of the leading order ChPT does not yield quantitatively different results~\cite{Ren:2014vea, Ren:2016aeo}.}

\begin{figure*}[t]
  \centering
 \includegraphics[width=8.0cm]{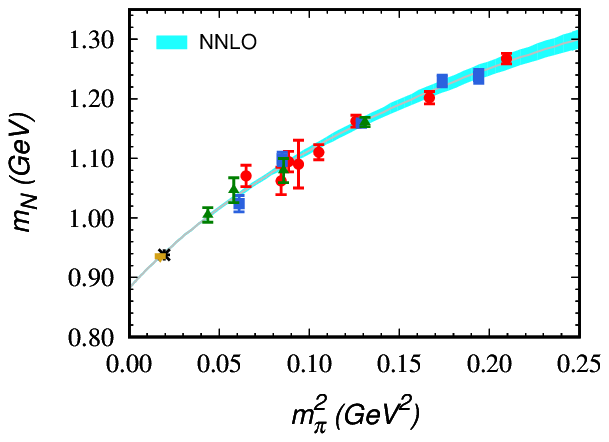} ~~~~
 \includegraphics[width=8.0cm]{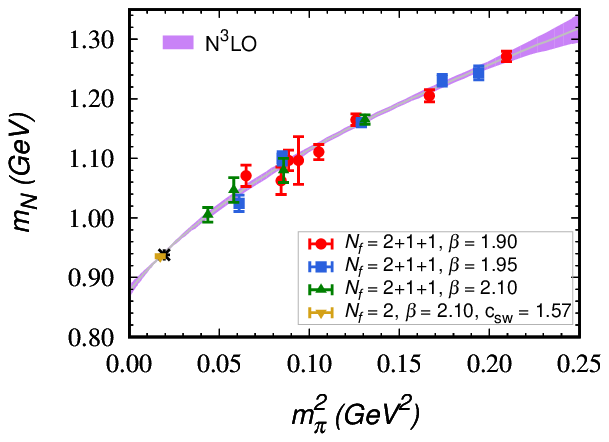}
 \caption{ETMC data with finite volume corrections subtracted in comparison with the best fits in the N$^2$LO and N$^3$LO BChPT. }
 \label{Fig:op34}
\end{figure*}

\section{Results and discussions}\

\subsection{Pion-nucleon sigma term from Feynman-Hellmann theorem}
In this subsection, we perform a least-square fit to the 18 sets of lattice QCD data (in the lattice unit) of Ref.~\cite{Alexandrou:2017xwd},  which are summarized in Table~\ref{Tab:LQCDdata}, together with the nucleon mass $m_N = 0.938$ GeV at the physical pion mass $m_\pi= 0.135$ GeV. 

At NNLO, one only has two LECs, namely $m_0$ and $c_1$, as shown in Eq.~(\ref{Eq:nucleonmass}).
On the other hand, the four lattice spacings $a$  should also be determined self-consistently,
according to Ref.~\cite{Alexandrou:2017xwd}. As a result, in total we have 6 free parameters to describe the ETMC nucleon masses. 
The best fit yields a $\chi^2/\mathrm{d.o.f}=0.87$, which is already smaller than $1.0$, contrary to the HB ChPT case~\cite{Alexandrou:2017xwd}. The values of $m_0$ and $c_1$ are tabulated in Table \ref{Tab:LECs-sigma}, and the four lattice spacings  are 
\begin{eqnarray}\label{Eq:latspacing}
  a_{N_f=2+1+1,\beta=1.90}&=&0.0964(12)~\mathrm{fm},\nonumber\\ 
  a_{N_f=2+1+1,\beta=1.95}&=&0.0855(9)~\mathrm{fm},\nonumber\\
  a_{N_f=2+1+1,\beta=2.10}&=&0.0661(7)~\mathrm{fm},\nonumber\\
  a_{N_f=2,\beta=2.10}&=&0.0933(3)~\mathrm{fm}.
\end{eqnarray}
We note that the so-determined lattice spacings are in good agreement with those determined
in the HB ChPT fit with the small scale expansion scheme up to ``N$^3$LO''~\cite{Alexandrou:2017xwd}.
In our studies up to N$^3$LO, there are three more LECs, namely $c_2$, $c_3$, and $\alpha$, resulting in a total of $9$ parameters. We note, however, that the ETMC data cannot  unambiguously fix the $5$ LECs and the lattice spacings simultaneously. As a result, we chose to fix the lattice spacings at the values determined at the NNLO.
The resulting fit is shown in Table~\ref{Tab:LECs-sigma}, and the description of the ETMC data is slightly improved in comparison with that of NNLO. 
One can see that the values of $m_0$ and $c_1$ are consistent with the ones from the $\mathcal{O}(p^3)$ fit with slightly larger uncertainties. We note that our $c_1$ is almost the same as that given in the studies of pion-nucleon scattering~\cite{Alarcon:2011zs,Yao:2016vbz,Hoferichter:2015tha,Siemens:2016jwj}. However, at N$^3$LO, the three LECs $c_2$, $c_3$, and $\alpha$ can not be determined precisely. Compared with those of Refs.~\cite{Alarcon:2011zs,Yao:2016vbz,Hoferichter:2015tha,Siemens:2016jwj}, the values of $c_2$ and $c_3$ are different, particularly, $c_2$ is negative. In addition, if the LECs $c_2$ and $c_3$ were fixed at those of Ref.~\cite{Hoferichter:2015tha}, the fit-$\chi^2/\mathrm{d.o.f.}$ would increase to more than one, but the corresponding pion-nucleon sigma term would not change much.

\begin{table*}[t]
\caption{ Fitted LECs of the $\mathcal{O}(p^3)$ and $\mathcal{O}(p^4)$  EOMS BChPT with the delta-isobar contribution, as well as the predicted $\sigma_{\pi N}$. The numbers in the parentheses are the
statistical uncertainties at the 68.3\% confidence level.}
\centering 
\begin{tabular}{c|c|ccccc|c}  
\hline\hline
  & $\chi^2/\mathrm{d.o.f.}$  &$m_0~(\mathrm{GeV})$&$c_1~(\mathrm{GeV}^{-1})$ & $\alpha~(\mathrm{GeV}^{-3})$  & $c_2~(\mathrm{GeV}^{-3})$ & $c_3~(\mathrm{GeV}^{-3})$ & $\sigma_{\pi N}~(\mathrm{MeV})$  \\
  \hline  
$\mathcal{O}(p^3)$ & 2.77 & $0.868\pm 0.002 $ & $ -1.17\pm 0.05 $ & -- & -- & -- & $59.7\pm 0.4$ \\
 $\mathcal{O}(p^4)$ & 0.78 &  $0.877\pm 0.010$  &  $-1.10 \pm 0.22 $ & $20.40\pm 12.07$ & $-8.79\pm 5.27$ & $-2.30\pm 1.78$ & $53.0\pm 6.8$  \\
\hline\hline
\end{tabular}
\label{Tab:Deltacon}
\end{table*}

In Fig. \ref{Fig:op34}, we show the pion mass dependence of the nucleon mass as predicted by the $\mathcal{O}(p^3)$ and $\mathcal{O}(p^4)$ BChPT. Clearly, the agreement with data in both cases are of the same quality. It should be noted that in plotting the lattice QCD data, finite volume corrections have been subtracted, which can reach as large as a few tens of MeV for lattice QCD simulations with large $m_\pi$ and small $m_\pi L$, such as set 4, 5, 8, 9, and 17 of Table  \ref{Tab:LQCDdata}.

Since the ETMC data can be well described with $\chi^2/\mathrm{d.o.f.}<1.0$ up to NNLO and N$^3$LO in  covariant BChPT, we take the result of $\mathcal{O}(p^3)$ as the central value,  $\sigma_{\pi N}=50.2(1.2)(2.0)$ MeV, where the first uncertainty is statistical and second is theoretical originated from chiral truncations. We could as well choose the N$^3$LO prediction as our central value and obtain
  $\sigma_{\pi N}=52.2(6.6)(2.0)$ MeV. In the present case, we prefer to take the NNLO prediction because the ETMC data do not constrain very well the LECs at N$^3$LO.

In Refs.~\cite{Alvarez-Ruso:2013fza,Ren:2013dzt,Yao:2016vbz,Siemens:2016jwj}, the virtual $\Delta(1232)$ was found to be able to improve the convergence of BChPT in certain cases. 
Thus, following Ref.~\cite{Alvarez-Ruso:2013fza}, we  take the contribution of the virtual $\Delta(1232)$ to the nucleon mass into account  up to N$^3$LO and study its effect on the description of the ETMC data and on the prediction of  the pion-nucleon sigma term.
 The pertinent LECs are fixed in the following way: $h_A=2.85$~\cite{Alvarez-Ruso:2013fza} and the mass splitting  $\delta=m_{\Delta0}-m_0=0.292$ GeV. At N$^3$LO, the value of $c_{\Delta 1}$ is fixed by fitting the NLO delta-isobar mass 
$m_\Delta =m_{\Delta 0}-4c_{\Delta 1} m_\pi^2$ to its physical value, yielding $c_{\Delta 1} = (m_0-0.942)/(4 m_\pi^2)$. 
We take the lattice spacings as given in Eq.~(\ref{Eq:latspacing}) and present the fitting results in Table~\ref{Tab:Deltacon}. 
At NNLO, including the $\Delta(1232)$ contribution increases the fit-$\chi^2/\mathrm{d.o.f.}$ to about $2.8$, similar to what happened in Refs.~\cite{Alvarez-Ruso:2013fza,Ren:2013dzt}. While, at $\mathcal{O}(p^4)$, the description of the lattice data is almost the same as that without the $\Delta(1232)$ contribution, and the obtained pion-nucleon sigma term is $\sigma_{\pi N}= 53.0(6.8)$ MeV, which agrees with the one obtained without the $\Delta(1232)$ contribution within uncertainties. 
One may conclude that the contribution of the $\Delta(1232)$ can be absorbed by the LECs of the nucleon only case up to N$^3$LO, consistent with the finding in the SU(3) study~\cite{Ren:2013dzt}.

One must note that we did not provide a comprehensive
assessment of theoretical uncertainties and they can be much
larger. On one hand, they could come from the use of the
SU(2) BChPT to study the $N_f = 2 + 1 + 1$ lattice QCD data
and from our neglect of the lattice spacing artifacts. On the
other hand, they can also come from our choice for the decay
constant $f_0$ and the renormalization scale $\mu$. For instance,
we noticed that instead of $f_0 = 0.0871$ GeV, the choice of
$f_0 = 0.0922$ GeV~\cite{Chin:2016} decreases the central value of $\sigma_{\pi N}$
by $2\sim3$ MeV at both NNLO and N$^3$LO. Nonetheless, such a
choice still yields a $\chi^2/\mathrm{d.o.f.} < 1$ and therefore cannot be
distinguished from our original choice.

\begin{figure*}[t]
  \centering
 \includegraphics[width=14cm]{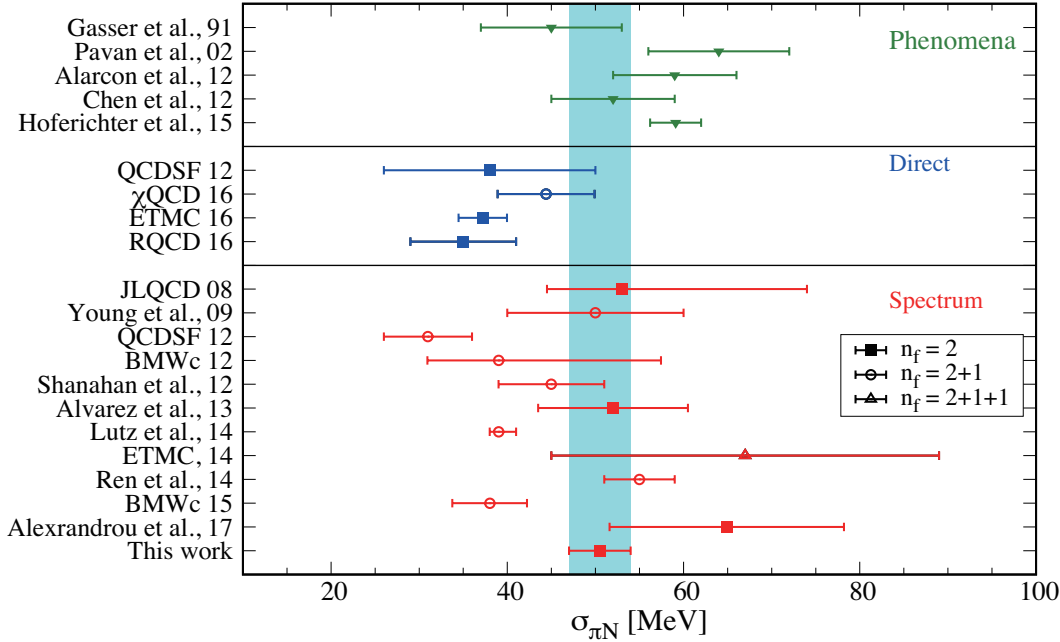}
 \caption{ Pion-nucleon sigma term, $\sigma_{\pi N}$, obtained in the phenomenological approaches, the lattice direct and spectrum methods, respectively. 
 The light-blue band represents the result obtained in the present work. 
 The rest data are taken from  \cite{Gasser:1990ce}~(Gasser {\it et al.}, 91), \cite{Pavan:2001wz}~(Pavan et al., 02), 
 \cite{Alarcon:2011zs}~(Alarcon {\it et al.} 12),  \cite{Chen:2012nx} ~(Chen {\it et al.} 12),
 \cite{Hoferichter:2015dsa}~(Hoferichter {\it et al.} 15) in the phenomenological studies, 
   \cite{Bali:2011ks}~(QCDSF 12), \cite{Yang:2015uis}~($\chi$QCD 16), 
   \cite{Abdel-Rehim:2016won}~(ETMC 16),
   \cite{Bali:2016lvx}~(RQCD 16) with the lattice direct method, 
\cite{Ohki:2008ff} (JLQCD 08), \cite{Young:2009zb}~(Young {\it et al.} 09),
    \cite{Horsley:2011wr}~(QCDSF 12), \cite{Durr:2011mp}~(BMWc 12), 
    \cite{Shanahan:2012wh}~(Shanahan  {\it et al.} 12), 
    \cite{Alvarez-Ruso:2013fza}~(Alvarez {\it et al.} 13), 
    \cite{Lutz:2014oxa}~(Lutz {\it et al.} 14), 
    \cite{Ren:2014vea}~(Ren {\it et al.}, 14),  
   \cite{Alexandrou:2014sha}~(ETMC 14), 
   \cite{Durr:2015dna}~(BMWc 15), 
  and \cite{Alexandrou:2017xwd}~(Alexandrou {\it et al.} 17) with the spectrum method.}
 \label{Fig:sigma}
\end{figure*}

A recent  study of the pion-nucleon scattering with the Roy-Steiner equations~\cite{Hoferichter:2015dsa} found that the effects of isospin breaking on the pion-nucleon sigma term is around $3$ MeV, which is comparable to the uncertainty from chiral truncations. Therefore, the isospin breaking effects on the $\sigma_{\pi N}$  should be carefully investigated. However, the ETMC data is obtained in the isospin limit and therefore cannot determine the
four LECs $c_5$, $f_1,~f_2,~f_3$, needed to parametrize the leading order isospin breaking between the $u$ and $d$ quarks~\cite{Hoferichter:2015hva}.~\footnote{Once $N_f=1+1+1$ lattice data become available, such as  those of the BMW collaboration, we can predict the pion-proton and pion-neutron sigma terms and  evaluate the isospin breaking effect.}

It is interesting to note that the central value of our predicted $\sigma_{\pi N}$  is smaller than the sigma term,  $64.9$ MeV, obtained in Ref.~\cite{Alexandrou:2017xwd}  by fitting to the same lattice QCD data. 
This difference  can be traced back to the fact that  HB ChPT can only describe the ETMC data with a $\chi^2/\mathrm{d.o.f.} \approx 1.6$ at
NNLO. Taking into account the uncertainty of chiral truncations, our result is consistent with the pion-nucleon sigma term,  $64.9(1.5)(13.2)$ MeV, of Ref.~\cite{Alexandrou:2017xwd}. Since at `` N$^3$LO'',  a $\chi^2/\mathrm{d.o.f.} \approx 1.1$  can be achieved~\cite{Alexandrou:2017xwd},  it is more reasonable to take the ``N$^3$LO'' prediction as the central value. In this case, one would obtain $\sigma_{\pi N} = 51.7(4.3) (13.2)$ MeV, whose central value is in better agreement with our result.

It should be noted that our predicted pion-nucleon sigma term is in  between that from the latest phenomenological studies, $\sigma_{\pi N}\sim 60$ MeV, and that from the recent LQCD calculations,  $\sigma_{\pi N}\sim 40$ MeV, as shown in Fig.~\ref{Fig:sigma}. 
In addition, the $\sigma_{\pi N}$ is consistent with
the value determined from our  three-flavor study~\cite{Ren:2014vea}. This is not surprising  since as shown in Ref.~\cite{Ren:2016aeo} the SU(3) and SU(2) BChPT are consistent with each other within uncertainties,
particularly for $m_\pi< 300$ MeV.  Second,
one should note that our results are tied to the quality of the lattice QCD data that we fitted. Nevertheless, our present study provides a further consistency check on the covariant BChPT we employed, which in many cases is essential to the determination of the $\sigma_{\pi N}$ via the spectrum method.

\subsection{Connection to  pion-nucleon scattering}

The LECs constants, $c_1,~c_2,~c_3$,  determined in the present study
can be used as inputs to perform a partial pion-nucleon scattering analysis and calculate 
the pion-sigma term with the Cheng-Dashen theorem and the scattering lengths. Such studies could provide a useful crosscheck on
the reliability of the determination of the pion-nucleon sigma term using the Feynman-Hellmann theorem. 

According to the Cheng-Dashen theorem~\cite{Cheng:1970mx}, the pion-nucleon sigma term reads
\begin{equation}
  \sigma_{\pi N} = \Sigma_d + \Delta_D - \Delta_{\sigma} - \Delta_{R},
\end{equation}
where $\Delta_{D}-\Delta_{\sigma}=(-1.8\pm 0.2)$ MeV~\cite{Hoferichter:2012wf}, $|\Delta_R| < 2 $ MeV~\cite{Bernard:1996nu}, 
$\Sigma_d = f_\pi^2 (d_{00}^+ + 2 m_\pi^2 d_{01}^+)$ with $d_{00}^+$ and $d_{01}^+$ the  sub-threshold parameters of pion-nucleon scattering.  Up to $\mathcal{O}(p^3)$~\cite{Becher:2001hv}~\footnote{Since there this no counter terms at $\mathcal{O}(p^3)$, the results of $d_{00}^+$ and $d_{01}^+$ in infrared ChPT are the same as the ones from the EOMS scheme.},   $d_{00}^+$ and $d_{01}^+$ are solely determined by $c_1$ and $c_3$ as
\begin{eqnarray}
  d_{00}^+ &= & -\frac{2m_\pi^2}{f_\pi^2} (2c_1-c_3) + \frac{g_A^2(3+8g_A^2)m_\pi^3}{64 \pi f_\pi^4}, \nonumber\\
   d_{01}^+ &=&  -\frac{c_3}{f_\pi^2} - \frac{g_A^2(77+48g_A^2)m_\pi}{768 \pi f_\pi^4}.
\end{eqnarray} 
With $c_1$ and $c_3$ in Table~\ref{Tab:LECs-sigma}, we obtain the sigma term as $45.6(2.2)$ MeV and  $51.8(2.2)$ MeV at NNLO and N$^3$LO, respectively. It is clear that these values are  consistent with the pion-nucleon sigma terms determined by
fitting to the  ETMC data within uncertainties. 

Recently,  Hoferichter {\it et al.}~\cite{Hoferichter:2015dsa} proposed a relationship between the pion-nucleon sigma term and the $S$-wave scattering lengths, $a^{1/2}$ and $a^{3/2}$, 
\begin{equation}
  \sigma_{\pi N} = (59.1\pm 3.1)~\mathrm{MeV} + \sum\limits_{I_s} c_{I_s} (a^{I_s} - \bar{a}^{I_s}),
\end{equation}
based on Roy-Steiner equations.  In Refs.~\cite{Hoferichter:2016ocj,RuizdeElvira:2017stg},  they showed that a small  $\sigma_{\pi N}$ is related to a even smaller value of   the $\pi N$ isoscalar scattering length, $a^+$. With our  $c_1,c_2,c_3$ tabulated in Table~\ref{Tab:LECs-sigma}, we obtain $a^+=-130.5\pm 195.1$ ($10^{-3} m_\pi^{-1}$), using the chiral expansions of Ref.~\cite{Yao:2016vbz}.  The central value is much smaller than  the one obtained from the pion-nucleon scattering analysis, $a^+=-14.8$  ($10^{-3} m_\pi^{-1}$)~\cite{Siemens:2016jwj}, but consistent within uncertainties.  Such a difference is partly
due to the negative    $c_2$  obtained in our study in comparison with the positive one from $\pi N$ scattering
and partly due to the fact that at $\mathcal{O}(p^4)$, we could not constrain well  $c_2$ and $c_3$ simply by fitting to the
ETMC nucleon masses, consistent with the finding of Ref.~\cite{Alvarez-Ruso:2013fza}.

\section{Summary }

We have reanalyzed the latest ETMC simulations of the nucleon mass and extracted the eagerly wanted pion-nucleon sigma term. We
showed that because of the use of the covariant baryon chiral perturbation theory, we were able to minimize theoretical uncertainties and obtain a pion-nucleon sigma term, 
$\sigma_{\pi N}=50.2(1.2)(2.0)$ MeV,
consistent with those determined from the $N_f=2+1$ and $N_f=2$ analyses, although more lattice QCD data
on the nucleon mass, and even on some complementary observables,
are still needed to further reduce theoretical uncertainties. 

With the LECs $c_1$, $c_2$, and $c_3$ determined by fitting to the ETMC nucleon masses, 
we also predicted the pion-nucleon sigma term using the Cheng-Dashen theorem and the scattering length $a^+$ of pion-nucleon scattering.
The pion-nucleon sigma term is consistent with that determined from the Feynman-Hellmann theorem, but the
scattering length only marginally agrees with the one from the phenomenological studies.

In order to better understand  the current tension between the pion-nucleon sigma terms from the lattice QCD calculations  
and those from the pion-nucleon scattering analyses and to better constraint the values of $c_2$ and $c_3$, 
a combined study of the lattice QCD nucleon masses and the pion-nucleon scattering data in
the same framework, such as the present one,  is in urgent need.

\begin{acknowledgments}
X.-L.R. thanks the valuable discussions with De-Liang Yao and Jun-Xu Lu about pion-nucleon scattering. 
This work was partly supported by the National Natural
Science Foundation of China (NSFC) under Grants 
No. 11735003, No. 11522539, 11375024,  and No. 11775099, by DFG
and NSFC through funds provided to the Sino-German CRC
110 “Symmetries and the Emergence of Structure in QCD”
(NSFC Grant No. 11621131001, DFG Grant No. TRR110),
the China Postdoctoral Science Foundation under Grants No.
2016M600845, No. 2017T100008, and the Fundamental Research
Funds for the Central Universities.

\end{acknowledgments}

\end{document}